\documentclass[aps,pra,floatfix,twocolumn,tightenlines,a4paper,superscriptaddress]{revtex4}

\usepackage{amsmath}
\usepackage{amsfonts}
\usepackage{graphicx}
\usepackage{epsfig}
\usepackage{color}
\usepackage{xspace}

\begin{document}
\title{Persistent Entanglement in the Classical Limit}
\author{M.J.~Everitt}
\email{m.j.everitt@physics.org}
\author{T.D.~Clark}
\email{t.d.clark@sussex.ac.uk}
\author{P.B.~Stiffell}
\affiliation{Centre for Physical Electronics and Quantum Technology, School of Science and
Technology, University of Sussex, Falmer, Brighton, BN1 9QT, U.K.}
\author{J.F.~Ralph}
\affiliation{Department of Electrical and Electronic Engineering, Liverpool University,
Brownlow Hill, Liverpool, L69 3GJ, U.K.}
\author{A.R.~Bulsara}
\affiliation{Space and Naval Warfare Systems Center, Code 2363, 53560 Hull Street,
San Diego, California 92152-5001, USA}
\author{C.J.~Harland}
\affiliation{Centre for Physical Electronics and Quantum Technology, School of Science and
Technology, University of Sussex, Falmer, Brighton, BN1 9QT, U.K.}

\begin{abstract}
\vspace*{10pt}
  The apparent  difficulty in recovering  classical nonlinear dynamics
  and chaos from standard quantum  mechanics has been the subject of a
  great deal of interest over  the last twenty years. For open quantum
  systems  -  those coupled  to  a  dissipative  environment and/or  a
  measurement  device -  it  has been  demonstrated that  chaotic-like
  behaviour can  be recovered in  the appropriate classical  limit. In
  this paper,  we investigate  the entanglement generated  between two
  nonlinear  oscillators,   coupled  to   each  other  and   to  their
  environment.   Entanglement  - the  inability  to factorise  coupled
  quantum  systems  into their  constituent  parts  -  is one  of  the
  defining features of quantum mechanics. Indeed, it underpins many of
  the recent  developments in quantum technologies. Here  we show that
  the entanglement characteristics  of two `classical' states (chaotic
  and periodic solutions) differ significantly in the classical limit.
  In particular,  we show that significant levels  of entanglement are
  preserved only in the chaotic-like solutions.
\end{abstract}
\maketitle

The correspondence principle is one of the fundamental building blocks
of modern physics.  Stated simply,  the principle insists that any new
theory should contain  the existing (experimentally verified) theories
in   some   appropriate   limit   or  approximation.    Without   this
correspondence,  it  would  not  be  possible to  build  new  physical
theories  without undermining  previous theory  and  experiment.  When
quantum mechanics  was originally proposed, the  natural expression of
the correspondence  principle could be  stated as \emph{`If  a quantum
  system  has a  classical analogue,  expectation values  of operators
  behave, in  the limit  $\hbar\rightarrow 0$, like  the corresponding
  classical  quantities'}  \cite{Mer98}. Whilst  this  works for  many
systems, problems sometimes arise when dealing with nonlinear systems
\cite{Cas79}. More  generally,  there are  often  found to  be
difficulties obtaining  nonlinear classical equations  from the linear
Schr\"{o}dinger equation.   This is a significant problem  and a great
deal  of work  has been  done to  analyse the  way in  which classical
nonlinearities express  themselves in isolated  (Hamiltonian) systems. 
In this  paper we  deal with an  associated problem,  namely nonlinear
behaviour of quantum systems that  are coupled to an environment (open
quantum systems).

Three main approaches exist for dealing with open quantum systems. The
first, based  on work by  Feynman and Vernon \cite{Fey63},  treats the
environment  as  a source  of  dissipation  and  decoherence and  then
calculates the average evolution  of the quantum system by integrating
out the environmental degrees of  freedom. This provides a good method
for large ensembles of quantum systems, but the average evolution does
not  reflect all  types of  nonlinear behaviour,  such as  chaos.  The
second  approach  is  to  examine  the evolution  of  the  probability
distribution of the quantum system in phase space, having averaged out
the environment \cite{Hab98}.  In this  case, it can be shown that the
quantum  evolution  of  the  probabilities  is  very  similar  to  the
evolution of the corresponding classical system, which is also coupled
to a noisy environment. The  third approach, often called {\it quantum
  trajectories}  \cite{Car93,Gis93,Gis93b,Heg93,Wis96,Ple98},  is  the
one adopted  here.  This method  uses continuous weak  measurements on
the environment  and results in modified quantum  evolution, which can
be described by a  stochastic Schr\"{o}dinger equation.  This equation
reduces  to the  previous cases  when averaged  over an  ensemble, but
provides different  trajectories for  individual systems. The  type of
trajectories depend  on the type  of measurement that is  performed on
the  environment  \cite{Wis96}.    Different  trajectories  are  often
referred to as {\it unravellings}  of the Master equation, which would
describe the  average quantum evolution. For  chaotic systems, quantum
trajectories show  behaviour that is  very similar to  their classical
counterparts   in  the   correspondence  limit   $\hbar\rightarrow  0$
\cite{Spi94,Bru96,Per98}. This   paper   demonstrates  that   the
chaotic-like quantum dynamics have other properties that do not reduce
to the classical  limit when $\hbar$ becomes small.   We show that the
entanglement  between two  coupled chaotic  oscillators 
can be significant  - even when  the trajectories
are essentially classical.

\begin{figure*}[!t]
  \begin{center}
    \resizebox*{0.9\textwidth}{!}{\includegraphics{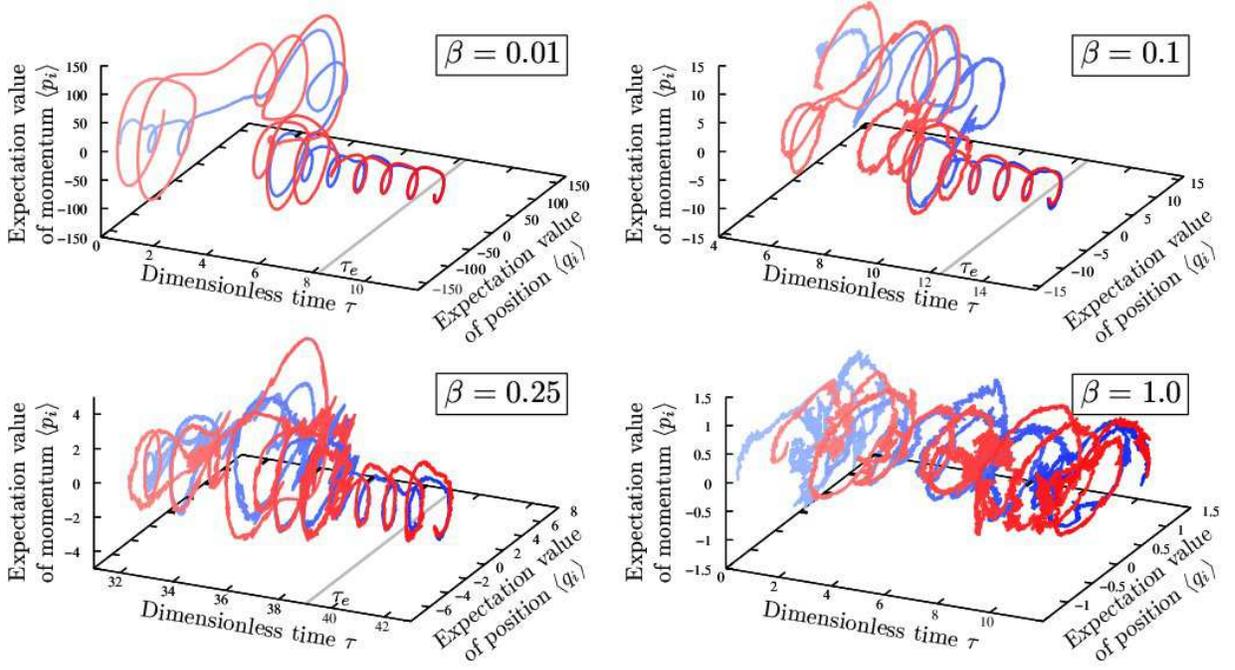}}
  \end{center}
  \caption{The dynamics of the expectation values of position and
    momentum as a  function of time (normalised to  drive periods) for
    $\beta=0.01$,  $\beta=0.1$,   $\beta=0.25$  and  $\beta=1.0$.  The
    dynamics have been  taken over the same duration  and displayed so
    that, with the exception  $\beta=1.0$, they approximately align at
    the time $(\tau_e)$ at which entrainment occurs.}
\end{figure*}
In this paper we have chosen to consider, as a convenient example, two
coupled  Duffing  oscillators  that  are described  by  Quantum  State
Diffusion  (QSD) \cite{Gis93,Gis93b}.  Duffing  oscillators were
chosen  because  they  are  a  standard example  of  classical  chaos,
describing   a  driven   oscillator  that   contains  a   third  order
nonlinearity. In addition single Duffing oscillators have been studied
extensively using  quantum trajectory models  \cite{Bru96,Per98}.  The
QSD unravelling corresponds to a unit-efficiency heterodyne measurement
(or ambi-quadrature  homodyne detection) on  the environmental degrees
of  freedom \cite{Wis96}. Similar chaotic-like  dynamics can  be
obtained from the Duffing  oscillator using most unravellings, and we
consider another unravelling ({\it quantum jumps} - corresponding to photon number
detection on the environment) below, and we show that similar entanglement properties can
be obtained. In QSD, the evolution of the  state 
vector $\left| \psi\right\rangle$ is given
by the increment (It\^{o} equation)\cite{Gis93,Gis93b,Heg93}
\begin{eqnarray}\label{eq:qsd}
\left\vert d\psi\right\rangle  &  =&-\frac{i}{\hbar}H\left\vert \psi
 \right\rangle dt\nonumber\\
&&  +\sum_{j}\left[  \left\langle L_{j}^{\dagger}\right\rangle L_{j}-\frac
{1}{2}L_{j}^{\dagger}L_{j}-\frac{1}{2}\left\langle L_{j}^{\dagger
}\right\rangle \Bigl\langle L_{j}\Bigr\rangle \right]  \left\vert
\psi\right\rangle dt\nonumber\\
&&  +\sum_{j}\left[  L_{j}-\left\langle L_{j}\right\rangle \right]  \left\vert
\psi\right\rangle d\xi
\end{eqnarray}
where  the  Linblad  operators   $L_{i}$  represent  coupling  to  the
environmental  degrees of  freedom, $dt$  is the  time  increment, and
$d\xi$     are     complex     Weiner     increments     such     that
$\overline{d\xi^2}=\overline{d\xi}=0$        and       $\overline{d\xi
  d\xi^{*}}=dt$ \cite{Gis93,Gis93b,Heg93}. The first term on the right
hand side  of (\ref{eq:qsd}) deals with  the Schr\"{o}dinger evolution
of the system  while the second (drift) and  third (fluctuation) terms
describe the decohering effects of the environment on the evolution of
the state vector.   To solve these equations can  be a computationally
demanding  problem  and  approaching  the  classical  limit  generally
requires the use of many basis states and extremely long run times.

In the classical limit, QSD  for one Duffing oscillator reproduces the
chaotic attractors (Poincar\'{e}  sections) of the classical equations
\cite{Bru96,Per98}. In this work, we extend this analysis and consider
two  identical, coupled  Duffing oscillators.  The  Hamiltonian (total
energy) for each oscillator is given by
\begin{equation}
H_{i}=\frac{1}{2}p_{i}^{2}+\frac{\beta^{2}}{4}q_{i}^{4}-\frac{1}{2}q_{i}
^{2}+\frac{g_{i}}{\beta}\cos\left(  t \right)  q_{i}+\frac
{\Gamma_{i}}{2}(q_{i}p_{i}+p_{i}q_{i})
\end{equation}
where $q_{i}$ and $p_{i}$ are  the position and momentum operators for
each   oscillator.    Here   the   Linblad\   operators   are   simply
$L_{i}=\sqrt{2\Gamma_{i}} a_{i}$  (for $i=1,2$), where  $a_{i}$ is the
oscillator  lowering operator.   Following \cite{Bru96,Per98},  we set
the parameters  for the drive and  the damping to  be $g_{i}=0.3$\ and
$\Gamma_{i}=0.125$  respectively.   The  Hamiltonian for  the  coupled
system then takes the form
\begin{equation}\label{eq:hamSys}
H=H_{1}+H_{2}+\mu q_{1}q_{2}
\end{equation}
where $\mu=0.2$ is  a measure of the strength  of the coupling between
the two oscillators.

Now, if we are to consider the correspondence principle in more detail
we see  that $\hbar\rightarrow 0$  is not the only  interpretation. By
this  we  mean that,  mathematically,  it  is  equivalent to  consider
$\hbar$ remaining fixed (as indeed  it does) and scale the Hamiltonian
so  that  the  relative  motion  of  the  expectation  values  of  the
observables   becomes   large   compared   with   the   minimum   area
$\left(\hbar/2\right)$  in the phase  space.  In  this paper,  this is
achieved by the introduction of a scaling parameter $\beta$, where the
system  behaves more  classically as  $\beta$ tends  to zero  from its
maximum value of one. The choice of $g_{i},\Gamma_{i}$ and the various
values  of $\beta$  come from  the previous  work on  the  QSD Duffing
oscillator and dissipative quantum chaos \cite{Bru96,Per98}.

We solve  the equation of  motion for the state  vector (\ref{eq:qsd})
using  the  Hamiltonian (\ref{eq:hamSys})  starting  with the  initial
states centred  on points taken  from the Poincar\'{e} section  of the
uncoupled classical  oscillators. Explicitly, this initial  state is a
tensor product of coherent states  for which the expectation values in
position  and momentum  are centred  in  $q-p$ phase  plane at  points
chosen from  the classical Poincar\'{e} section. In  Figure~1, we show
the evolution of  the expectation values of position  and momentum for
the two  coupled oscillators (in blue  and red) as a  function of time
for  a selection  of  values of  $\beta$.   From these  figures it  is
apparent that the initial phase space trajectory of the coupled system
is, for low $\beta$ - the classical limit, very similar to the that of
the classical  Duffing oscillator.  In  this small $\beta$  limit, the
uncertainties  in position  and  momentum are  a  very small  fraction
(typically  a few percent)  of the  oscillations, indicating  that the
dynamics are  essentially classical.   As $\beta$ increases  to unity,
this  structure becomes  harder to  resolve and  finally in  the fully
quantum limit  $(\beta=1)$ it is indiscernible. However,  it is clear,
for $\beta=0.01, 0.1$ and $0.25$, that after a certain time period the
chaotic  behaviour of the  coupled oscillators  gives way  to regular,
almost  periodic motion.   The oscillators  can be  said to  have {\it
  entrained},  i.e. their  nonlinear oscillations  are  stabilised and
synchronised by  the coupling. We  note for future reference  that for
$\beta=1$, it is impossible to  determine the presence of a chaotic or
periodic attractor. Also,  it is not possible to  determine whether or
not there is any level  of entrainment between these oscillators. This
is because, in the quantum limit, the quantum noise is large enough to
drive  the system  away  from  periodic attractors  and  to swamp  the
fractal structure found in the chaotic classical attractor.

\begin{figure}[!th]
\begin{center}
\resizebox*{0.4\textwidth}{!}{\includegraphics{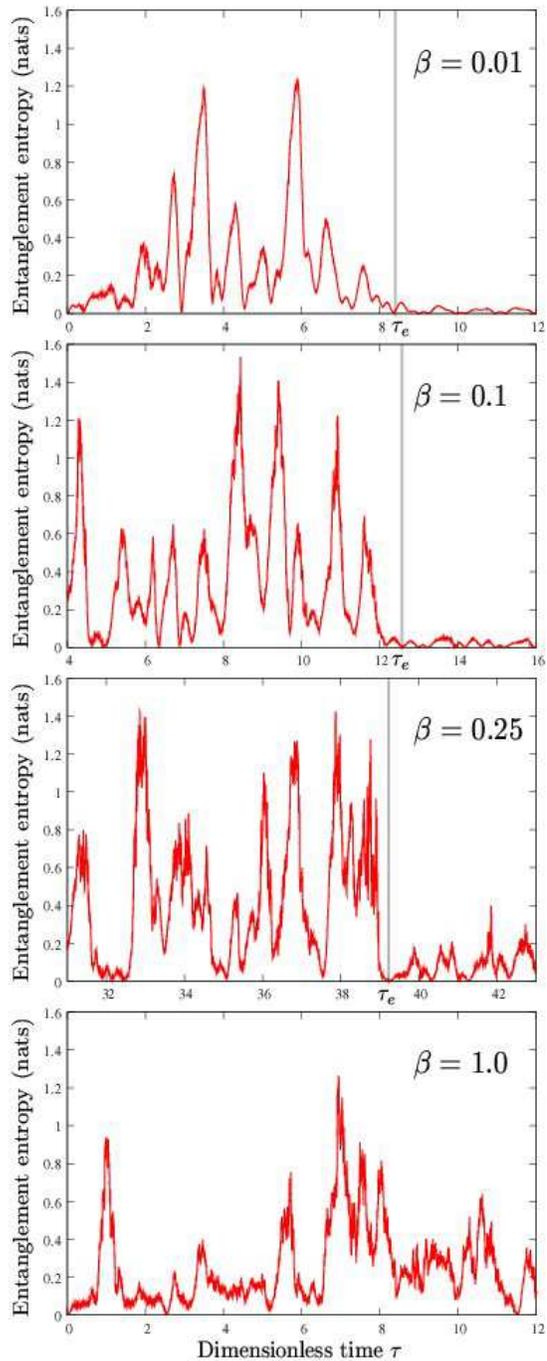}}
\end{center}
\caption{Entanglement entropy as a function of time (normalised to drive periods) for  $\beta=0.01$,
  $\beta=0.1$,  $\beta=0.25$  and   $\beta=1.0$.  By  comparison  with
  Figure~1  we can see  that there  is a  sudden drop  in entanglement
  entropy  after  the time  $\tau_e$  at  which  the oscillators  have
  entrained.}
\end{figure}

We note  that, when the  two oscillators are entrained,  the classical
correlations   between  the   oscillators  are   at  a   maximum  (the
oscillations are, after all, fairly well synchronised). It is natural,
therefore, to ask what level  of quantum correlation is present within
this system and  how the chaotic and periodic  phases of motion affect
this correlation.   We measure  this quantum correlation  by computing
the  entanglement entropy  (mutual  information) for  the system  from
\cite{Nie00}
\begin{equation}
E=S\left(  \rho_{1}\right)  +S\left(  \rho_{2}\right)  -S\left(  \rho\right)
\end{equation}
where $\rho=\left\vert  \psi\right\rangle \left\langle \psi\right\vert
$\    is     the    density    operator    of     the    system    and
$\rho_{i}=\operatorname*{Tr}_{j\neq  i}[\rho]$ is the  reduced density
operator    for   each   oscillator.    Here   $S\left(\rho_{i}\right)
=\operatorname*{Tr}[\rho_{i}\ln\rho_{i}]$  is the von  Neumann entropy
\cite{Nie00}.  As  QSD models  the evolution  of  a pure
state, $S\left(\rho\right)=0$\  at all times. Since  entanglement is a
purely  quantum mechanical phenomenon,  one might  expect that  when a
quantum  system approaches  the  classical limit  the two  oscillators
would not be strongly entangled.

In Figure~2 we show the entanglement entropy as a function of time for
the four  different values  of $\beta$ as  used in Figure~1,  over the
same  corresponding   time  intervals.  By   comparing  Figure~2  with
Figure~1, we can see that its entanglement remains relatively large as
long  as the  system undergoes  chaotic motion.  It is  only  when the
system becomes  entrained that the entanglement  drops approaching the
classical limit.   As we have  already said, the uncertainties  in the
$\beta  = 0.01$  case are  very small  when compared  to  the system's
evolution in phase  space. In this case, the  system is well localised
in phase space. The correspondence  principle would seem to imply that
these objects  are behaving  classically even though  the entanglement
(in  the chaotic  phase) does  not seem  to be  diminishing.  Once the
oscillators  entrain, the system  becomes approximately  separable and
the  entanglement entropy  falls  as $\beta$  reduces.  These  results
would  indicate that  quantum oscillators  operating in  an apparently
classical regime can generate  significant levels of entanglement, but
only in  chaotic-like states.  It is worth  mentioning that  the close
connection between chaos in  quantum systems and entanglement has been
noted by  Wang et al.  for Hamiltonian systems \cite{Wan03},  and they
have recently considered the behaviour in the classical limit \cite{Gho04}. 
However, the system discussed in reference \cite{Gho04} is non-chaotic
and only entanglement between the system and its environment is considered.
Here, we consider the entanglement between two systems (each of which is
coupled to an environment).

\begin{figure}[h]
\begin{center}
\resizebox*{0.48\textwidth}{!}{\includegraphics{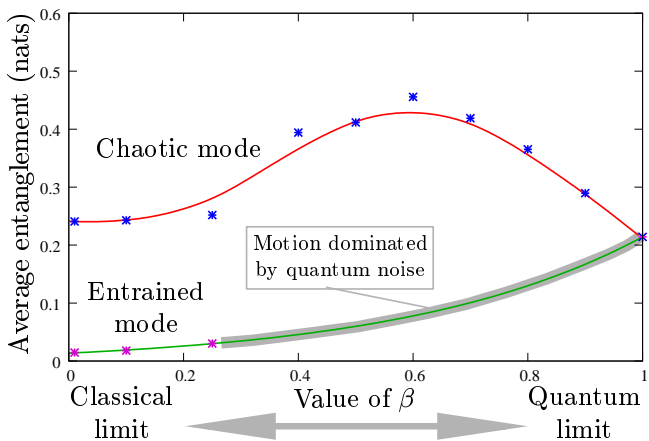}}
\end{center}
\caption{Mean entanglement entropy as a function of $\beta$ for the chaotic-like
  and periodic  (entrained) states. Here we see  that the entanglement
  entropy for system  in the chaotic state does  not vanish as $\beta$
  approaches the classical regime. \label{fig:average.qsd}}
\end{figure}

Each solution  computed using  quantum state diffusion  corresponds to
one realisation or experiment. We  obtain a more general picture if we
compute the  average entanglement as  a function of $\beta$  over many
experiments: the  mean entanglement entropy  \cite{Wan03}, rather than
the  entanglement of  the  average  (i.e. mixed)  state.  We show  the
results of  such a computation in  Figure~3. The  mean entanglement in
the chaotic  state does not  appear to vanish as  $\beta\rightarrow 0$
and,  possibly more  surprisingly, its  maximum  is not  at $\beta=1$.  
However,  we also  see  that  when this  system  entrains (at  smaller
$\beta$ values) the entanglement is rapidly suppressed. Figure~3 also
shows lines indicating  the best fit to the mean  data points.  In the
case of  the entrained  state, the best  fit is an  exponential, which
includes  the  quantum limit  ($\beta=1$),  even  though  there is  no
evidence for an entrained state for this $\beta$-value.

For  the preceding  results,  we have  used the quantum state  diffusion
unravelling of  the master equation. This  unravelling makes assumptions
about the underlying measurement process which is applied to the
environment: unit efficiency  heterodyne detection for QSD. However, 
studies have shown that the level of entanglement is dependent upon which
unravelling is used \cite{Nha04}. It is natural then to ask whether the
properties of the entanglement that we have described above are particular
to the QSD unravelling. To consider this, we choose another unravelling -
the quantum jumps unravelling \cite{Car93,Heg93}.
This model is very different from QSD as it is based on a discontinuous 
photon counting measurement process, rather than a diffusive continuous evolution. 
However,  it  has  already  been  shown  that  this  rather  different
unravelling can produce, in  the classical limit, the expected chaotic
dynamics of  the Duffing oscillator~\cite{brun97}.   It seems natural,
therefore,  to use  the quantum  jumps unravelling as  a second  test
unravelling  for our  example system.   In this  model, the  pure state
stochastic evolution equation for quantum jumps is given by
\begin{eqnarray}
  \label{eq:jumps}
  \left| d \psi \right\rangle
  & = & - \frac{i}{\hbar}H   \left|  \psi \right\rangle dt    \nonumber\\
  &   & - \frac{1}{2}\sum_j \left[L_j^\dag L_j - 
          \left\langle L_j^\dag L_j \right\rangle \right] 
          \left|  \psi \right\rangle dt \nonumber\\
  &   & + \sum_j \left[ \frac{L_j}{\sqrt{\left\langle L_j^\dag L_j \right\rangle}} 
        - 1 \right] \left|  \psi \right\rangle dN_j
\end{eqnarray}
where  $dN_j$   is  a  Poissonian   noise  process  such   that  $dN_j
dN_k=\delta{jk}dN_j$,  $dN_j  dt=0$ and  $\overline{dN_j}=\left\langle
  L_j^\dag L_j \right\rangle  dt$, i.e. jumps occur  randomly at a
rate that is determined by $\left\langle L_j^\dag L_j \right\rangle$.

\begin{figure}[h]
\begin{center}
\resizebox*{0.48\textwidth}{!}{\includegraphics{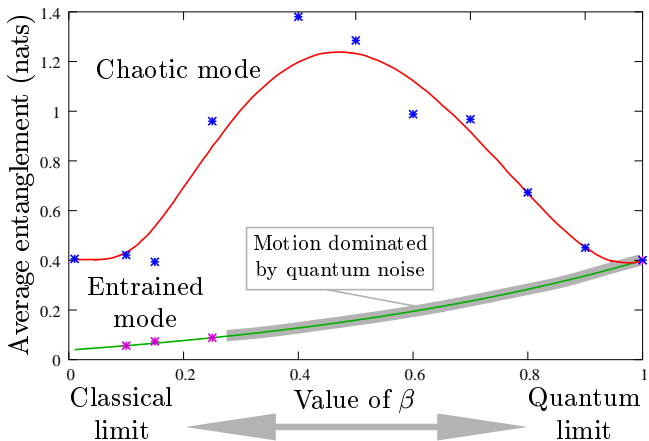}}
\end{center}
\caption{The calculation  of figure~\ref{fig:average.qsd} using quantum 
  jumps instead  of quantum  state diffusion. Again  we show  the mean
  entanglement entropy  as a function of $\beta$  for the chaotic-like
  and periodic (entrained) states.  As with quantum state diffusion we
  see  that when  using  quantum jumps  the  entanglement entropy  for
  system in  the chaotic state  does not vanish as  $\beta$ approaches
  the classical regime.\label{fig:jumps}}
\end{figure}
As  we   can  see   from  figure~\ref{fig:jumps}  the   quantum  jumps
unravelling produces entanglement that persists in the classical limit
so long as the oscillators  remain in the chaotic-like solution.  Once
again, if the  oscillators are seen to entrain,  then the entanglement
entropy  drops  significantly.   The  dependence of  the  entanglement
entropy  on $\beta$  is similar  in  form for  both unravellings  (one
continuous  and one  discontinuous).  We  find, as  might  be expected
\cite{Nha04}, that the entanglement  entropy for quantum jumps differs
from  that of  QSD. However,  even  though the  average level  differs
between unravellings,  the behaviour of the average  entanglement - in
both chaotic-like and  non-chaotic - as a function  of $\beta$ is very
similar.  We note that the data points shown in figure~\ref{fig:jumps}
exhibit  a   slightly  more  erratic   nature  than  those   shown  in
figure~\ref{fig:average.qsd}.   The explanation for  this lies  in the
fact that the the entanglement entropies obtained for the evolution of
the system under the the quantum jumps unravelling take much more time
to settle to a good average than for the QSD unravelling. Without very
long computational times (e.g.   several months) we cannot achieve much
more  significant  accuracy  with  the computational  power  currently
available to us.

In this  paper we have considered  the classical limit  of two coupled
nonlinear quantum  oscillators.  The analysis  is based on standard
examples from classical chaos and open quantum systems.  We have shown
that when these oscillators  enter the almost periodic entrained state
the entanglement falls rapidly  as the system approaches the classical
regime,  as  expected.  Conversely,  we  have  also  shown that  these
oscillators can remain significantly  entangled even when the dynamics
of  the system  appear  to be  classical,  although only  when in  the
chaotic state.  Moreover, by using both the quantum state diffusion and
quantum jumps models, we have  shown that these results are not unique
to  a particular  unravelling of  the  master equation.   We find  the
results  presented in  this paper  are surprising,  since  the quantum
correlations  represented by  the  non-zero entanglement  do not  have
classical   counterparts   yet   they   persist  in   the   classical
(correspondence) limit.

\begin{acknowledgments}
  We would like to thank the EPSRC for its support of this work though
  its Quantum Circuits  Network. The authors would also  like to thank
  T.P.~Spiller and  W.~Munro for  interesting  and
  informative  discussions.    MJE  would  also  like   to  thank  
  P.M.~Birch for his helpful advice.
\end{acknowledgments}

\bibliography{references}

\end{document}